\documentclass[%
 reprint,
 amsmath,amssymb,
 aps,
]{revtex4-1}
\usepackage{color}
\usepackage{epsf}
\usepackage{graphicx}
\usepackage{dcolumn}
\usepackage{bm}
\usepackage{dsfont}
\usepackage{appendix}

\usepackage{color}
\usepackage{ulem}

\begin{document}

\preprint{APS/123-QED}

\title{Quark and gluon condensates in QCD reevaluated }

\author{Renata Jora
	$^{\it \bf a}$~\footnote[1]{Email:
		rjora@theory.nipne.ro}}

\affiliation{$^{\bf \it a}$ National Institute of Physics and Nuclear Engineering PO Box MG-6, Bucharest-Magurele, Romania}

\begin{abstract}
We compute the quark and gluon condensates in $QCD$ with $N$ colors and $N_f$ flavors based on the renormalization group equations and on the knowledge of a single scale $\Lambda_{QCD}$. For $N=3$ and $N_f=3$ our findings are in the good range for $0.2\leq\Lambda_{QCD}\leq 0.36$ and in excellent agreement with the results in the literature from sum rules for a value $\Lambda_{QCD}=0.28$.

\end{abstract}
\maketitle

Quark and gluon condensates are vacuum expectations values of local operators. They are often considered in the context of quark-hadron dualities and calculated as such. However the general view is that these condensates represent characteristic scales of the QCD vacuum that may be regarded as independent of the hadron degrees of freedom \cite{Brodsky}.

The quark condensate in QCD always indicates dynamical supersymmetry of the chiral symmetry $SU(3)_L\times SU(3)_R$ which is associated with the presence of Nambu-Goldstone bosons. It is expected that the chiral group is broken down to $SU(3)_V$ such that all quark condensates are at least in first order the same.

For the case when one restricts the chiral group to $SU(2)_L\times SU(2)_R$ the relation between the quark condensates and the pions becomes evident in the context of the Gell-Mann-Oakes-Renner relation \cite{G1}, \cite{G2}:
\begin{eqnarray}
f_{\pi}^2m_{\pi}^2=-2(m_u+m_d)\langle \bar{q}q\rangle.
\label{relgell87}
\end{eqnarray}
Here $m_{\pi}$ is the mass of the pion, $f_{\pi}$ is the pion decay constant, $m_u$ and $m_d$ are the current quark masses  and $\langle\bar{q}q\rangle$ is the vacuum quark condensate. Further relations of interest may be obtained from the current algebra \cite{current} and the axial vector Ward-Takahashi identity.

Although there is less information regarding the gluon condensate both quark and gluon condensates have been computed in the literature in the lattice QCD approach \cite{lattice}, sum rules \cite{sum1}- \cite{Jamin} and chiral models \cite{chiral}.

The purpose of the present work is to compute once again the quark and gluon condensates from the renormalization group equations of the quantum field theory of QCD without reference to the hadron structure and properties.
Although usually in the literature the formation of the quark and gluon condensates are intertwined with confinement here we will consider them unrelated topics. We are less interested in finding the exact point in the space $(N,N_f)$ where the condensates form and the actual phase transitions take place. Instead we will aim to determine the values of these condensates for a specific value of the coupling constant in terms of the intrinsic scale of the theory $\Lambda_{QCD}$, which is considered known.

We consider QCD with $N$ colors and $N_f$ flavors. We work with the all order beta function inspired by the supersymmetric NSVZ beta function  proposed in \cite{Sannino}:
\begin{eqnarray}
\beta(g)=\frac{ \partial g}{\partial \ln(\mu)}=-\frac{g^3}{16\pi^2}\frac{\frac{11}{3}N-\frac{N_f}{3}(2+\Delta_R\gamma_R)}{1-\frac{g^2}{8\pi^2}\frac{17}{11}N},
\label{betfunc64554}
\end{eqnarray}
where,
\begin{eqnarray}
\Delta_R=1+\frac{17}{11}\frac{2N^2}{N^2-1},
\label{delta66454}
\end{eqnarray}
and,
\begin{eqnarray}
\gamma_r=-\frac{d\ln(m)}{d\ln(\mu)}=3\frac{N^2-1}{N}\frac{g^2}{16\pi^2}.
\label{res54663}
\end{eqnarray}
We will also need the anomalous dimension of the fermions wave function which is given by:
\begin{eqnarray}
\gamma_1=\frac{g^2}{12\pi^2}.
\label{res6453442}
\end{eqnarray}

We first consider the quark two point correlator in the Fourier space which we will denote by $Q(p,M)=\frac{i}{p^{\mu}\gamma_{\mu}}f(\frac{p^{\rho}\gamma_{\rho}}{M})$ where $f$ is an arbitrary function that we need to determine. We apply the renormalization group equations:
\begin{eqnarray}
\Bigg[M\frac{\partial}{\partial M}+\beta(g)\frac{\partial}{\partial g}+2\gamma_1\Bigg]Q(p^{\rho}\gamma_{\rho},M)]=0.
\label{renorm65677489}
\end{eqnarray}
Here  $M$ is the renormalization scale. We pick a specific point where $\beta(g)=\infty$ and consequently $(1-\frac{g_1^2}{8\pi^2}\frac{17}{11}N)=0$. Then either,
\begin{eqnarray}
\frac{\partial}{\partial g}Q(p,M)=0,
\label{part4665}
\end{eqnarray}
or $Q(p,M)$ has a complicated dependence on $g$. We assume that the choice in Eq. (\ref{part4665}) is valid. We solve  Eq. (\ref{renorm65677489}) to find:
\begin{eqnarray}
Q_1={\rm Tr}\Bigg[Q(p,M)\Bigg]={\rm Tr}\Bigg[\frac{i}{p^{\mu}\gamma_{\mu}}a[\frac{p^{\rho}\gamma_{\rho}}{M}]^{2\gamma_1}\Bigg],
\label{sol7888}
\end{eqnarray}
where $a$ is  a constant determined as $a=1$ from the renormalization condition $Q(p,M)|_{p^{\rho}\gamma_{\rho}=M}=\frac{i}{p^{\mu}\gamma_{\mu}}$.

The quark condensate can be then calculated as:
\begin{eqnarray}
&&\langle \bar{q}q\rangle =v=\int \frac{d^4 p}{(2\pi)^4}Q_1=
\nonumber\\
&&i\int \frac{d^4 p_E}{(2\pi)^4}{\rm Tr}\Bigg[Q(p_E,M)\Bigg]\approx
\nonumber\\
&&-\frac{3}{4\pi^2}\int d(p_E^2) p_E^2\frac{1}{M^{2\gamma_1}}(p_E^2)^{\gamma_1-\frac{1}{2}}=
\nonumber\\
&&-\frac{3}{4\pi^2}\frac{2}{2\gamma_1+3}M^3.
\label{finalexprqu}
\end{eqnarray}
Here we integrated up to the scale $M$ which corresponds to the coupling constant $g_1$ and the integral was performed in the euclideean space.

The next step is to consider the two point correlator in the Fourier space for the gluon field $G(p^2,M^2)=-ig^{\mu\nu}\frac{1}{p^2}h(\frac{p^2}{M^2})$ (we work in the Feynman gauge) where again $h$ is  an arbitrary function, solution of the  associated renormalization group equation:
\begin{eqnarray}
\Bigg[\frac{\partial}{\partial M}+\beta(g)\frac{\partial}{\partial g}+2\gamma_A\Bigg]G(p^2,M^2)=0.
\label{resgl788}
\end{eqnarray}
Here $\gamma_A$ is the anomalous dimension of the gluon wave function:
\begin{eqnarray}
\gamma_A=-\frac{3g^2}{16\pi^2}.
\label{gammaa43552}
\end{eqnarray}
We consider the same fixed coupling constant $g_1$ and by the same arguments as for the quark condensate we find:
\begin{eqnarray}
G_1={\rm Tr} \Bigg[G(p^2,M^2)\Bigg]=b4(N^2-1)\frac{-i}{p^2}(\frac{p^2}{M^2})^{\gamma_A}.
\label{solgl867759}
\end{eqnarray}
The constant b is calculated as before from the renormalization condition at scale $M$ as $b=1$.
The gluon condensate can then be computed approximately from:
\begin{eqnarray}
&&T=\langle G^a_{\mu\nu}G^{a\mu\nu}\alpha\rangle=\int \frac{d^4p}{(2\pi)^4}2p^2 G_1\frac{g_1^2}{4\pi}=
\nonumber\\
&&\int \frac{d^4p_E}{(2\pi)^4}8(N^2-1)(\frac{p_E^2}{M^2})^{\gamma_A}=
\nonumber\\
&&\frac{1}{2\pi^2}\frac{1}{\gamma_A+2}M^4(N^2-1)\frac{g_1^2}{4\pi}.
\label{finalexpr74665}
\end{eqnarray}
Here $\alpha=\frac{g_1^2}{4\pi}$.

The scale $M$ where the coupling constant is $g_1=(\frac{88\pi^2}{51})^{\frac{1}{2}}$ is calculated by integrating Eq. (\ref{betfunc64554}) from $\Lambda_{QCD}$ to $M$.

In this paper we determined the quark and gluon condensate based on first principle and on the knowledge of a single parameter $\Lambda_{QCD}$. In order to do that we needed  a particular value of the coupling constant at some scale compatible with the beta function. We used instead of the standard QCD beta function computed at five loops in \cite{Vermaseren} the all beta function inspired by the NSVZ beta function proposed by Pica and Sannino in \cite{Sannino}.  Our choice had double purpose:  first it simplified the calculations in a great measure, second it made it easier to extrapolate our findings to the supersymmetric QCD similar calculations of the condensates. The latter analysis will be reserved for future work.

We considered standard QCD with $N=3$ and $N_f=3$.

\begin{figure}
\begin{center}
\epsfxsize = 10cm
\epsfbox{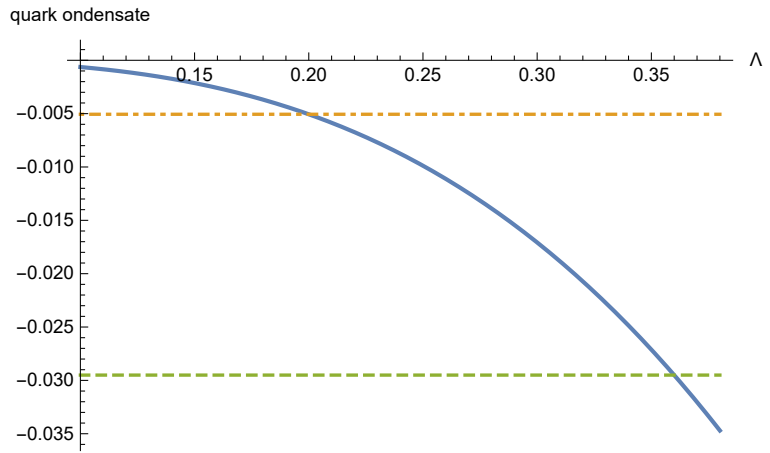}
\end{center}
\caption[]{%
Plot of the quark condensate $\langle \bar{q}q\rangle=v$ in terms of $\Lambda$, the intrinsic scale of the theory. The intersection of the thick line with the dotdashed line indicates the value of the condensate for $\Lambda=0.2$ whereas the intersection with the dashed line indicates the value of the condensate for $\Lambda=0.36$..
}
\label{quark}
\end{figure}

\begin{figure}
\begin{center}
\epsfxsize = 10cm
\epsfbox{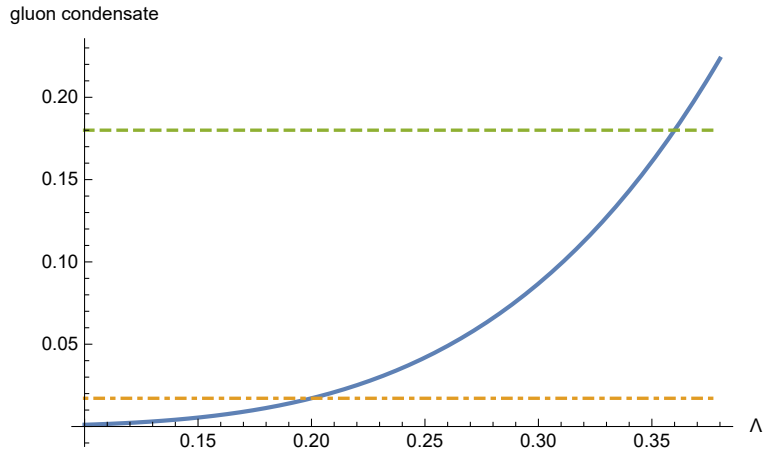}
\end{center}
\caption[]{%
Plot of the gluon  condensate $\langle G^a_{\mu\nu}G^{a\mu\nu}\alpha\rangle=T$ in terms of $\Lambda$, the intrinsic scale of the theory. The intersection of the thick line with the dotdashed line indicates the value of the condensate for $\Lambda=0.2$ whereas the intersection with the dashed line indicates the value of the condensate for $\Lambda=0.36$.
}
\label{gluon}
\end{figure}

 In Fig. (\ref{quark}) we plotted  the quark condensate in terms of the intrinsic scale of the theory $\Lambda$. For the range of $0.2 \leq\Lambda\leq 0.36$ $GeV$ we obtained  $-0.05\leq v\leq -0.03$ $GeV^3$. In Fig. (\ref{gluon}) we plotted the gluon condensate in terms of the intrinsic scale of the theory $\Lambda$. For the range of $0.2 \leq\Lambda\leq 0.36$ $GeV$ we got  $0.017\leq T\leq 0.18$ $GeV^4$.

The gluon condensate was extracted for non-perturbative QCD from charmonium sum rules in \cite{SVZ}:
\begin{eqnarray}
\langle \alpha_sG^2\rangle\approx 0.04\,\, GeV^4.
\label{svstr665}
\end{eqnarray}
In the subsequent years many other calculations were performed, some of them close to the value in Eq. (\ref{svstr665}), others like in lattice QCD for example \cite{lattice} with a larger range of values.

The light quark condensate has also been estimated in sum rules as \cite{Jamin}:
\begin{eqnarray}
\langle \bar{q}q\rangle =-(267\pm \,\,16 MeV)^3,
\label{jamin5666}
\end{eqnarray}
or lattice QCD \cite{lattice}:
\begin{eqnarray}
\langle \bar{q}q\rangle =-(283\pm 2\,\, MeV)^3.
\label{lattice}
\end{eqnarray}

It might be more useful to find specific values of the condensates based on our knowledge of  low energy QCD.  For that we use the estimate for the gluon condensate in the sum rules approach as taken from \cite{sum3}  $\langle \frac{g^2}{4\pi}G^2\rangle\approx 0.07$ $GeV^4$  to determine $\Lambda_{QCD}=0.28$ GeV and  with this value we computed the quark condensate $\langle\bar{q}q\rangle\approx -0.015$ $GeV^3=-(0.244 \,\,GeV)^3$. This compares very well with other results in the literature computed using different other methods \cite{lattice}-\cite{chiral}.

The method can be used for determining the quark and gluon condensate for an arbitrary number of flavors and colors pending on the knowledge of the characteristic scale of the theory $\Lambda$.


\begin{thebibliography}{30}


\bibitem{Brodsky} S. J. Brodsky, C. D. Roberts, R. Shrook, P.C. Tandy, Phys. Rev. C {\bf 82}, 022201 (2010).
\bibitem{G1} M. Gell-Mann, R. J. Oakes and B. Renner, Phys. Rev. {\bf 175}, 2195 (1968).
\bibitem{G2} R. F. Dashen, Phys. Rev. {\bf 183}, 1245 (1969).
\bibitem{current} S. L. Adler and R. F. Dashen, "Current Algebras and Applications to Particle Physics", Benjamin, New York 1968.
\bibitem{lattice} C. McNeile, A. Bazavov, C. T. H. Davies, R. J. Dowdall, K. Hornbostel, G. P Lepage, H. D. Trottier, Phys. Rev. D {\bf 87}, 034503 (2013); C. J. Morningstar and M. J. Peardon, Phys. Rev. D {\bf 60}, 034509 (1999); W. J. Lee and D.Weingarten, Phys. Rev. D {\bf 61}, 014015 (2000); G. S. Bali et al., Phys. Lett. B {\bf 309}, 379  (1993); C. McNeile et al [UKQCD Collaboration], Phys. Rev. D {\bf 63}, 114503 (2001); L. C. Gui et al.[CLQCD COllaboration], Phys. Rev. Lett. {\bf 110}, no. 2, 021601 (2013); G. Burgio, F. Di Renzo, G. Marchesini and E. Onofri, Phys. Lett. B {\bf 422}, 219 (1998); R. Horley, P. E. L. Rakow and G. Schierholz, Nucl Phys. (Proc. Sup.) B {\bf 106}, 870 (2002); M. D"Elia, A. Di Giacomo and E. Meggiolaro, Phys. Lett. B {\bf 408}, 315 (1997).
\bibitem{sum1} C. A. Dominguez, M. Kremer, N. A. Papadopoulos and K. Schilcher, Z. Phys. C {\bf 27}, no. 3, 481-489 (1985).
\bibitem{sum2} M. A. Shifman, A. I. Vainshtein and V. I. Zakharov, Nucl. Phys. B {\bf 147}, 385 ,448 (1979); S. Narison and G. Veneziano, Int. J. Mod. Phys. A {\bf 4}, 2751 (1981); S. Narison, Z. Phys. C {\bf 26}, 209 (1984).
\bibitem{sum3} S. Narison, Phys. Lett. B {\bf 706}, 412-422 (2012).
\bibitem{SVZ} M. A. Shifman, A. I. Vainshtein and V. I. Zakharov, Nucl. Phys. B {\bf 147}, 385, 448 (1979).
\bibitem{Ioffe} B. I. Ioffe and K. N. Zyablyuk, Eur. Phy. J. C {\bf 27}, 229 (2003); B. I. Ioffe, Progr. Part. Nucl. Phys. {\bf 56}, 232 (2006).
\bibitem{Jamin} M. Jamin, Phys. Lett. B {\bf 538}, 71-76 (2002).
\bibitem{chiral} R. Williams, C. S. Fischer and M.R. Pennington, Phys. Lett. B {\bf 645}, 167-172 (2007); M. Albaladejo and. J. A. Oller, Phys. Rev. Lett. {\bf 101}, 252002 (2008); F. Giacosa, T. Gutsche, V. E. Lyubovitskij and A. Faessler, Phys. Rev. D. {\bf 72}, 094006 (2005); M. Chanowitz, Phys. Rev. Lett. {\bf 95}, 172001 (2005); A. H. Fariborz, Int. J. Mod. Phys. a {\bf 19}, 2095 (2004); A. H. Fariborz, Phys. Rev. D {\bf 77}, 054030 (2006); C. Amsler and F. E. Close, Phys. Rev. D {\bf 53}, 295 (1996).
\bibitem{Vermaseren} F. Herzog, B. Ruijl, T. Ueda, J. A. M. Vermaseren, A. Vogt, JHEP {\bf 02},090 (2017).
\bibitem{Sannino} C. Pica and F. Sannino, Phys. Rev. D {\bf 83}, 116001 (2011).

\end{thebibliography}
\end{document}